\documentclass[
 reprint,
 superscriptaddress,
 amsmath,
 amssymb,
 aps,
 pra,
]{revtex4-2}

\usepackage[dvipdfmx]{graphicx}
\usepackage{bm}
\usepackage{color}

\begin{document}

\preprint{APS/123-QED}
\title{Gaussian-wave-packet model for single-photon generation based on cavity quantum electrodynamics under adiabatic and nonadiabatic conditions}

\author{Takeru Utsugi}
\email{utsugitakeru@akane.waseda.jp}
\affiliation{Department of Applied Physics, Waseda University, 3-4-1 Okubo, Shinjuku, Tokyo 169-8555, Japan}
\author{Akihisa Goban}
\affiliation{Department of Applied Physics, Waseda University, 3-4-1 Okubo, Shinjuku, Tokyo 169-8555, Japan}
\author{Yuuki Tokunaga}
\affiliation{Computer \& Data Science Laboratories, NTT Corporation, Musashino 180-8585, Japan}
\author{Hayato Goto}
\affiliation{Corporate Research \& Development Center, Toshiba Corporation, Kawasaki, Kanagawa 212-8582, Japan}
\author{Takao Aoki}
\affiliation{Department of Applied Physics, Waseda University, 3-4-1 Okubo, Shinjuku, Tokyo 169-8555, Japan}

\date{\today}

\begin{abstract} 
For single-photon generation based on cavity quantum electrodynamics, we investigate a practical model assuming a Gaussian wavepacket. This model makes it possible to comprehensively analyze the temporal dynamics of an atom-cavity system with both adiabatic and nonadiabatic conditions using analytical expressions. These results enable us to clarify the relationship between pulse width and maximum success probability, over the full range of coupling regimes.
We demonstrate how to achieve a high success probability while keeping a short pulse width by optimizing the cavity transmittance parameter and the time-controlled exexternal field.
Our formulations provide a practical tool for efficient single-photon generation in a wide variety of experimental platforms.
\end{abstract}
\pacs{Valid PACS appear here}
\maketitle

\section{\label{intro}Introduction} 
Single-photon generation based on cavity quantum electrodynamics (QED) of three-level atoms is useful for photon-based quantum information processing (QIP) systems, due to the ability to control the photon wavepacket~\cite{vasilev2010single} and achieve the high success probability~\cite{goto2019figure}. 
For construction of high-speed QIP systems, it is desirable to minimize the output pulse width of the photon, while maintaining this high success probability.

Previously, the conditions of the output pulse width $\tau$ for achieving a high success probability have been studied in two coupling regimes: the Purcell (bad cavity) and strong-coupling regimes~\cite{kuhn2010cavity}, which are determined by the coupling constant $g$, cavity field decay rate $\kappa$, and atomic polarization decay rate $\gamma$.
In the Purcell regime (${\kappa \gg g^{2}/\kappa \gg \gamma}$)~\cite{purcell1946,law1997deterministic}, photons are generated by cavity-enhanced Raman scattering, where the condition for high success probability is ${\tau\gg\kappa/g^{2}}$.
On the other hand, in the strong-coupling regime [${g \gg (\kappa, \gamma)}$]~\cite{kuhn1999controlled,vasilev2010single}, photons are generated by vacuum stimulated Raman adiabatic passage (vSTIRAP), so that the above condition becomes ${\tau\gg1/\kappa}$.
The same results have also been shown for single-photon storage, corresponding to time-reversal of single-photon generation, in both the Purcell~\cite{Gorshkov} and strong-coupling regimes~\cite{Dilley}.

Furthermore, the analytical expressions on the upper bound of success probability have been studied~\cite{vasilev2010single,goto2019figure}. In the Ref.~\cite{vasilev2010single}, a universal approach for generating the desired wavepacket shape has been proposed, and the photon generation efficiencies of various pulse shape have been discussed in the strong coupling regime. Recently, we have derived the upper bound on the success probability by optimizing the cavity external loss rate under the adiabatic condition~\cite{goto2019figure}. 
While recent work has investigated the success probability of nonadiabatic single-photon storage by numerical simulations of quantum master equations~\cite{Giannelli, Macha}, analytical expressions of the upper bound in the nonadiabatic condition are yet to be clarified.
In order to help design and optimize cavity QED experiments, analytical expressions for the success probability are desirable for both adiabatic and nonadiabatic conditions. Moreover, we wish to extend this analysis to be valid not only in the Purcell and strong-coupling regimes, but also in the weak-coupling [${(\gamma,\kappa)>g}$] and intermediate ($g\sim\kappa$) regimes.

In this paper, we investigate single-photon generation under both adiabatic and nonadiabatic conditions by introducing a Gaussian wavepacket model.
Assuming a Gaussian output pulse, we analytically formulate the upper bound of the success probability not only for arbitrary pulse width, but also for any coupling regime.
In addition, we show the arbitrary detuning can be compensated by controlling the external field and investigate the population dynamics and required external fields for any coupling regime.
We then provide a method for achieving a high success probability with a short pulse width by optimizing the cavity transmittance parameter.
Our formulations will be useful for engineering high-speed QIP systems, having high success probability and short pulse width, in a wide variety of experimental platforms.

This paper is organized as follows.
In Sec.~\ref{sec2}, we introduce a model of a cavity QED system, formulate the upper bound of the success probability, and discuss the reduction factor and pulse width dependence of the success probability.
In Sec.~\ref{sec3}, we provide a temporal analysis to show the dependence on the pulse width and the different coupling regimes. 
In Sec.~\ref{sec4}, we optimize the external loss rate of a cavity in both adiabatic and nonadiabatic conditions.
In Sec.~\ref{sec5}, we assume a typical experimental setup and provide a method for achieving a short pulse width while maintaining a high success probability by optimizing the cavity parameters.
Finally, we summarize our findings in Sec.~\ref{summary}.

\section{\label{sec2}Model}
In this section, we investigate a model assuming a Gaussian wavepacket for single-photon generation, and derive the upper bound of the success probability for both adiabatic and nonadiabatic conditions.

\begin{figure}
	\includegraphics[clip,width=8cm]{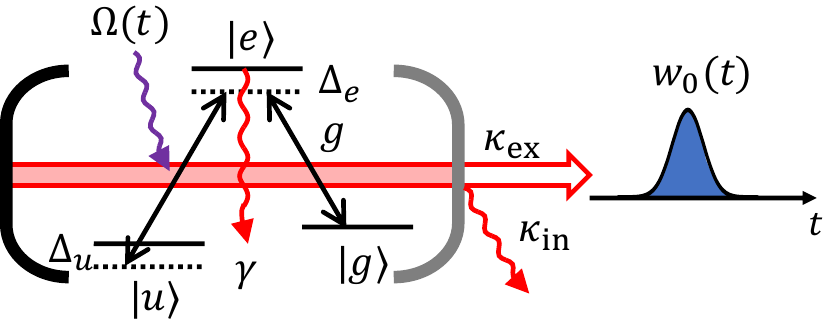}
	\caption{Cavity QED system for single-photon generation. The atom is initially prepared in the uncoupled state $|u\rangle$. $\kappa_{\mathrm{in}}$ and $\kappa_{\mathrm{ex}}$ are the cavity's internal and external loss rates (field decay rates), respectively, $g$ is the atom-cavity coupling rate via the $|g\rangle\leftrightarrow|e\rangle$ transition, $\Omega(t)$ is the Rabi frequency (including phase) of the time-controlled external field for the $|u\rangle\leftrightarrow|e\rangle$ transition, $\Delta_e$ and $\Delta_u$ are one-photon and two-photon detunings, respectively, and $\gamma$ is the atomic polarization decay rate due to spontaneous emission. $w_{0}(t)$ is the temporal waveform of the emitted photon's wavepacket, which we consider to be Gaussian shaped in this work.}
	\label{fig-system}
\end{figure}

Figure~\ref{fig-system} shows a schematic of a cavity QED system~\cite{vasilev2010single,goto2019figure}, where a $\Lambda$-type three-level atom is trapped in a one-sided cavity. The atom is initially in state $|u\rangle$, and the $|u\rangle\leftrightarrow|e\rangle$ transition is driven with an external classical field of Rabi frequency $\Omega(t)$, which includes the phase of the field with respect to the dipole moment of the trapped atom. The $|g\rangle\leftrightarrow|e\rangle$ transition is coupled to the cavity, and $w_{0}(t)$ is the temporal waveform of the emitted photon wavepacket. The total cavity field decay rate is $\kappa=\kappa_{\mathrm{in}}+\kappa_{\mathrm{ex}}$, where $\kappa_{\rm{in}}$ and $\kappa_{\rm{ex}}$ are the internal and external decay rates, respectively. Note that the atom-cavity coupling constant $g$, the atomic polarization decay rate $\gamma$, $\kappa_{\rm{in}}$, and $\kappa_{\rm{ex}}$ are time-independent parameters. 
Note also that since the photon generation by the reexcitation process, where the atom is reexcited after the inner spontaneous emissions from $|e\rangle$ to $|u\rangle$, is not useful for the typical QIP systems~\cite{goto2019figure}, we treat the inner spontaneous emissions as a loss.

For this system, we outline the derivation of the success probability based on previous works~\cite{vasilev2010single,goto2019figure}.
We describe the time evolution of the system by the following non-Hermitian Schr\"{o}dinger equation as ($c = \hbar = 1$)
\begin{gather}
	i|\dot{\psi}\rangle=\mathcal{H}|{\psi}\rangle,
	\label{eq-001}
	\\
	\intertext{where}
	|\psi \rangle 
	=
	\alpha_u |u \rangle |0 \rangle_c |0 \rangle + \alpha_e |e \rangle |0 \rangle_c |0 \rangle + \alpha_g |g \rangle |1 \rangle_c |0 \rangle
	\notag
	\\
	+\int_{-\infty}^{\infty}\alpha_k |g\rangle |0\rangle_c |k \rangle dk,
	\label{eq-002}
	\\
	\mathcal{H}
	=
	\left( \Delta_e-i\gamma \right) \sigma_{e,e} + \Delta_u \sigma_{u,u}
	+\int^{\infty}_{-\infty} kb^{\dagger}(k)b(k)dk
	\notag
	\\
	+ 
	i \left(\Omega\sigma_{e,u} - \Omega^{*}\sigma_{u,e} \right)
	+
	ig \left( a \sigma_{e,g} - a^{\dagger} \sigma_{g,e} \right)
	-i \kappa_{\mathrm{in}} a^{\dagger} a
	\notag
	\\
	+i \sqrt{  \frac{\kappa_{\mathrm{ex}}}{\pi}}\int^{\infty}_{-\infty}[b^{\dagger}(k)a-a^{\dagger}b(k)]dk.
	\label{eq-003}
\end{gather}

\noindent
In Eq.~(\ref{eq-001}), $|\psi\rangle$ is the state vector of the system, where the dot denotes the time derivative.
The total ket vector given by Eq.~(\ref{eq-002}) consists of the tensor products of three vectors, corresponding to an atom, and a photon inside and outside the cavity, where $\alpha_{j}$ ($j\in u, e, g, k$) are the complex amplitudes of the corresponding states.
In Eq.~(\ref{eq-003}), $a$ and $a^{\dagger}$ are the annihilation and creation operators for cavity photons, respectively, $b(k)$ and $b^{\dagger}(k)$ are the annihilation and creation operators for output-mode photons, respectively, with photon wavenumber, $k$,
where $|k\rangle=b^{\dagger}(k)|0\rangle$.
$\sigma_{m,l}=|m\rangle \langle l|$ ($m,l\in u, e, g$) are atomic operators.
Accordingly, the Schr\"odinger equation becomes
\begin{align}
\dot{\alpha_u}
=&
-i\Delta_u \alpha_u
-\Omega^{*} \alpha_e,
\label{eq-alpha-u}
\\
\dot{\alpha_e}
=&
-(\gamma + i \Delta_e) \alpha_e + \Omega \alpha_u + g \alpha_g,
\label{eq-alpha-e}
\\
\dot{\alpha_g}
=&
-\kappa \alpha_g - g \alpha_e,
\label{eq-alpha-g}
\\
\alpha_k
=&
\sqrt{\frac{\kappa_{\mathrm{ex}}}{\pi}}\int_{-\infty}^t \alpha_g(t')e^{-ik(t-t')}dt',
\label{eq-alpha-k}
\end{align}

\noindent
where we have used ${\int_{-\infty}^{t}\alpha_{g}(t')\delta(t-t')dt'=\alpha_{g}(t)/2}$~\cite{walls2007quantum}.

We derive the upper bound of the overall success probability $P_{S}$ by solving the differential equations (4)--(7) for given cavity QED parameters and pulse shape.
$P_{S}$ is defined by 
\begin{equation}
{P_{S}\equiv2\kappa_{\rm{ex}}\int_{t_{\rm{st}}}^{t_{\rm{end}}}|\alpha_{g}(t')|^{2}dt'},
\label{PSdef}
\end{equation}
where $t_{\rm{st}}$ and $t_{\rm{end}}$ are the start and end times of the photon emission, respectively.

First, we define the normalized waveform of the photon $w_0(t)$ by
\begin{gather}
w_0(t) 
=
\sqrt{\frac{2\kappa_{\mathrm{ex}}}{P_{S}}}\alpha_g(t),
\label{eq-w0}
\end{gather}
\noindent
where the intensity $|w_0(t)|^{2}$ is normalized in the range of ${t_{\rm{st}}<t<t_{\rm{end}}}$ as $\int_{t_{\rm{st}}}^{t_{\rm{end}}} |w_{0}(t)|^{2}dt = 1$.
The temporal waveform of the photons can be controlled by $\Omega(t)$~\cite{vasilev2010single,khanbekyan2017cavity}. 
In this work, we consider a Gaussian wavepacket with a width of $\tau>0$.
The temporal waveform is defined as 
\begin{align}
w_0(t)=\sqrt{\frac{1}{\sqrt{\pi}\tau}} \mathrm{exp}\left( - \frac{t^{2}}{2\tau^{2}}\right),
\label{Gaussian}
\end{align}
where photon emission starts at ${t_{\rm{st}}=-\infty}$ and ends at $t_{\rm{end}}=\infty$ while the pulse width of $\tau$ is a finite value. One potential waveform for constructing high-fidelity QIP systems is the Gaussian waveform, due to its tolerance against the effects of a temporal mode mismatch~\cite{rohde}. We note that the following discussion is also valid for the case of a truncated pulse shape, as discussed in Appendix C. 

Given the Gaussian waveform $w_{0}(t)$, we first study the temporal evolution of atomic states by solving the differential equations for ${\rho_{jj}(t)\equiv|\alpha_{j}(t)|^{2}~(j\in{u,e,g})}$.
From Eqs.~(\ref{eq-w0}) and (\ref{Gaussian}), $\rho_{gg}$ is obtained as
\begin{align}
\rho_{gg}(t)&=\frac{P_{S}}{2\sqrt{\pi}\kappa_{\rm{ex}}\tau}\mathrm{exp}\left( - \frac{t^{2}}{\tau^{2}}\right).
\label{eq-rhogg}
\end{align}
Substituting $\alpha_{g}$ obtained from Eqs.~(\ref{eq-w0}) and (\ref{Gaussian}) into Eq.~(\ref{eq-alpha-g}), $\rho_{ee}$ is also obtained as
\begin{align}
\rho_{ee}(t)&=\frac{P_{S}}{2\sqrt{\pi}\kappa_{\rm{ex}}g^{2}\tau}\left(\kappa-\frac{t}{\tau^{2}}\right)^{2}\mathrm{exp}\left( - \frac{t^{2}}{\tau^{2}}\right).
\label{eq-rhoee}
\end{align}

Furthermore, from the norm of $|\psi\rangle$ and Eqs.~(\ref{eq-alpha-u})--(\ref{eq-alpha-g}), we describe $\rho_{uu}$ as~\cite{vasilev2010single}
\begin{align}
\rho_{uu}(t)
=&
1-\rho_{ee}(t)-\rho_{gg}(t)
\notag\\
&-2\int^{t}_{t_{\rm{st}}}dt'\left[\gamma\rho_{ee}(t')+\kappa\rho_{gg}(t') \right].
\label{eq-ro1}
\end{align} 
Then, substituting Eqs.~(\ref{eq-rhogg}) and (\ref{eq-rhoee}) into Eq.~(\ref{eq-ro1}), we obtain $\rho_{uu}$ as 
\begin{multline}
\rho_{uu}(t) = 1-\frac{P_{S}\left[\gamma+2\kappa\tau^{2}(g^{2}+\gamma\kappa)\right]\left[\mathrm{Erf}\left(\frac{t}{\tau}\right)+1\right]}{4g^{2}\kappa_{\rm{ex}}\tau^{2}}\\
-\frac{P_{S}e^{ - \frac{t^{2}}{\tau^{2}}}}{2\sqrt{\pi}\kappa_{\rm{ex}}g^{2}\tau^{5}}\left\{t^{2}-t(\gamma+2\kappa)\tau^{2}+\left[g^{2}+\kappa(2\gamma+\kappa)\right]\tau^{4}\right\}.\\
\label{eq-rhouu}
\end{multline}

Here, we consider a physical requirement as follows,
\begin{gather}
\rho_{uu}(t)\ge0 \text{ for all } t\in [t_{\rm{st}},t_{\rm{end}}].
\label{init-cond2}
\end{gather}

Now, we derive the upper bound of $P_{S}$ given the cavity QED parameters ${(g,\gamma,\kappa_{\rm{in}},\kappa_{\rm{ex}})}$ and pulse width $\tau$ by using the physical requirement shown in Eq.~(\ref{init-cond2}). 
We examine the minimum value of $\rho_{uu}(t)$ using 
\begin{align}
\dot{\rho}_{uu}(t)&=\frac{P_{S}e^{-\frac{t^{2}}{\tau^{2}}}}{g^{2}\kappa_{\rm{ex}}\tau^{7}\sqrt{\pi}}(t-t_{0})(t-t_{+})(t-t_{-})=0,
\label{eq-tpm0}
\end{align}
where
\begin{align}
t_{0}&=\kappa\tau^{2},
\label{eq-t0}\\
t_{\pm}&=\frac{\kappa+\gamma}{2}\tau^{2}\pm\tau\sqrt{1-\tau^{2}\left[g^{2}-\left(\frac{\kappa-\gamma}{2}\right)^{2}\right]}.
\label{eq-tpm}
\end{align}
From ${\rho_{uu}(t_{\rm{st}})=1}$ and ${0\le\rho_{uu}\le1}$, the minimum value of $\rho_{uu}$ is obtained by one of these three points, i.e., ${\rho_{uu}(t_{m})\le\rho_{uu}(t)}$ for all $t$, where ${t_{m}\in\{t_{0}, t_{+}, t_{-}\}}$ is the time when $\rho_{uu}$ reaches the minimum value.

Therefore, the upper bound of $P_{S}$ for the given ${(g,\gamma,\kappa_{\rm{in}},\kappa_{\rm{ex}},\tau)}$ can be obtained by imposing the inequality of $\rho_{uu}(t_{m})\ge0$ on Eq.~(\ref{eq-rhouu}) and solving for $P_{S}$ as
\begin{widetext}
\begin{equation}
P_{S}^{\rm{max}}\equiv\min_{t_{m}\in\{t_{0}, t_{+}, t_{-}\}}\frac{\kappa_{\rm{ex}}}{\kappa}\frac{2C}{\left( 2C+1+\frac{1}{2\kappa^{2}\tau^{2}}\right)\left( \frac{\mathrm{Erf}(\frac{t_{m}}{\tau})+1}{2}\right)+\frac{e^{-\left(\frac{t_{m}}{\tau}\right)^{2}}}{2\sqrt{\pi}\kappa^{2}\gamma\tau^{3}}\left[ \left(\frac{t_{m}}{\tau^{2}}-\gamma-2\kappa\right)t_{m}+\kappa^{2}\tau^{2}+2\kappa\gamma\tau^{2}(C+1)\right]},
\label{Ps-gen}
\end{equation}
\end{widetext}
where $P_{S}^{\rm{max}}$ is the upper bound of $P_{S}$ in the finite pulse width, and $C\equiv g^{2}/(2\kappa\gamma)$ is a cooperativity parameter.
$P_{S}<P_{S}^{\rm{max}}$ occurs when the population of the uncoupled state $|u\rangle$ remains as $\rho_{uu}(t_{m})\neq0$.
Equation~(\ref{Ps-gen}) shows the relation between maximum $P_{S}$ and arbitrary cavity QED parameters and $\tau$. 
Here, we note that $P_{S}^{\rm{max}}$ is independent of detunings $\Delta_{u}$ and $\Delta_{e}$, and this generation efficiency can be achieved in the resonant case, or nonresonant case where optimally controlling the external field in the manner described in Appendix A. In other words, the effect of detunings can be canceled by controlling the amplitude and phase of the external field, and $P_{S}^{\rm{max}}$ can be achieved while emitting the Gaussian pulse at any detunings.

In the long pulse limit ($\tau\rightarrow\infty$), the upper bound of $P_{S}$ is given by substituting Eqs.~(\ref{eq-t0}) and (\ref{eq-tpm}) into Eq.~(\ref{Ps-gen}), and letting ${\tau\rightarrow\infty}$ as
\begin{align}
P_{S}^{\rm{max}}(\tau\rightarrow\infty) = \eta_{\rm{esc}} \frac{2C}{2C+1},
\label{caseii}
\end{align}
where ${\eta_{\rm{esc}}\equiv \kappa_{\rm{ex}}/\kappa}$ is the escape efficiency of the cavity. 
Indeed, Eq.~(\ref{caseii}) is consistent with previously known upper bound~\cite{vasilev2010single,goto2019figure,reiserer2015cavity}.

For finite $\tau$, temporal control of the output waveform forces a finite population to the uncoupled state, resulting in a reduction of $P_{S}$. Specifically, it can be shown that (see Appendix B)
\begin{align}
P_{S} \approx \left[1-\rho_{uu}(t=\infty)\right] P_{S}^{\rm{max}}(\tau\rightarrow\infty),
\label{eq-ano}
\end{align}
which is valid for $\tau\gg\frac{1}{\kappa\sqrt{2(2C+1)}}$. 
It can be seen from Eqs.~(\ref{caseii}-\ref{eq-ano}) that $\rho_{uu}(t=\infty)\rightarrow 0$ for $\tau \rightarrow \infty$.
Furthermore, for the case of $C \rightarrow \infty$, it can be shown that (see Appendix B)
\begin{align}
{P_{S}} = \eta_{\rm{esc}} [1-\rho_{uu}(t=\infty)],
\label{casei}
\end{align}
for arbitrary $\tau$. 

Thus, we can understand that the reduction of $P_{S} $ from unity is determined by three factors: $\eta_{\rm{esc}}$, $\rho_{uu}(t=\infty)$, and $C$.

\begin{figure}
	\includegraphics[clip,width=8cm]{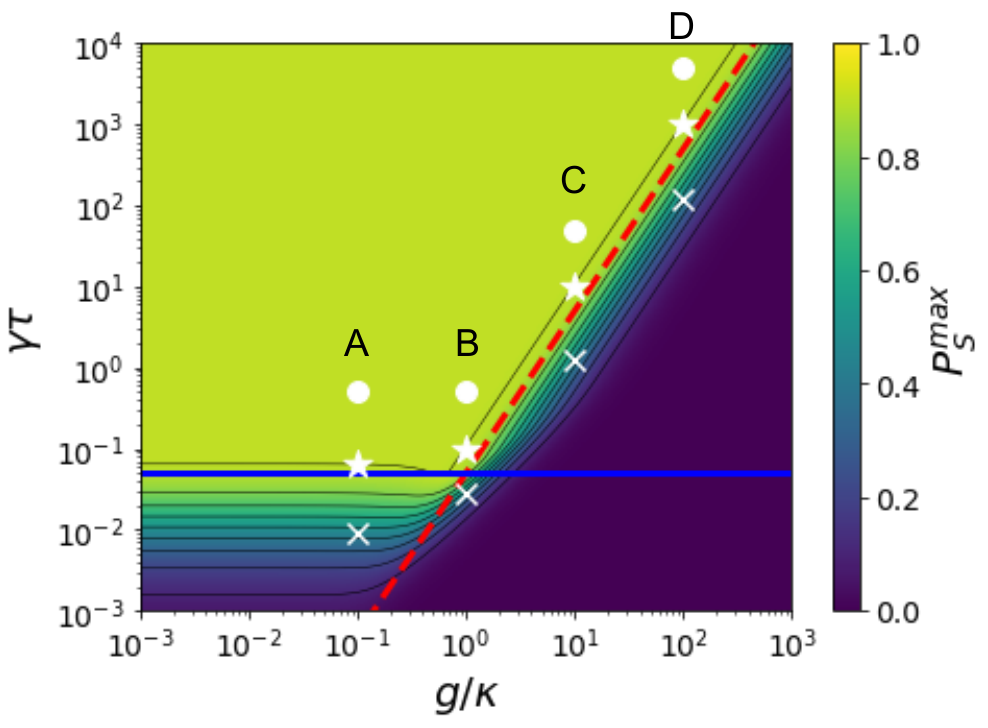}
	\caption{$P_{S}^{\rm{max}}$ as a function of $\gamma\tau$ and $g/\kappa$, with $C=10$ and $\eta_{\rm{esc}}=0.95$. The red dashed and blue lines represent $\tau=1/\kappa$ and $\kappa/g^{2}$, respectively. The circle, star and cross marks of A--D are the points analyzed in Figs. 3, 4, and 5, respectively.}
	\label{figPs}
\end{figure}

Figure~\ref{figPs} illustrates $P_{S}^{\rm{max}}$ as a function of $\gamma\tau$ and $g/\kappa$ given by Eq.~(\ref{Ps-gen}), where we set $C=10$ and $\eta_{\rm{esc}}=0.95$~\cite{memo}. 
The condition on the pulse width while maintaining high $P_{S}^{\rm{max}}$ depends on the coupling regimes, characterized by $g/\kappa$. From the red dashed and blue lines in Fig.~\ref{figPs}, we can determine the condition realizing high $P_{S}^{\rm{max}}$ when $C>1$ as
\begin{equation}
\tau \gg
\left\{
\begin{array}{ll}
	1/\kappa & \text{if } g>(\kappa,\gamma) \text{ or } \gamma>g>\kappa,\\
	\kappa/g^{2} & \text{if } \kappa>g^{2}/\kappa>\gamma,
\end{array}
\label{tauccase}
\right.
\end{equation}
where the first case represents two regimes: the strong-coupling regime [${g>(\kappa,\gamma)}$], and the weak-coupling regime under the condition of $C>1$ (${\gamma>g>\kappa}$). The second case represents the Purcell regime. 
These conditions are consistent with the bandwidth of the dressed atom in each regime.
This result is also consistent with the numerical analysis of single-photon storage under nonadiabatic conditions shown in~\cite{Giannelli}.
From Eq.~(\ref{tauccase}), we introduce a critical pulse width as 
\begin{equation}
\tau_{c} \equiv \mathrm{max}\left(\frac{1}{\kappa},~\frac{\kappa}{g^{2}}\right).
\label{tauc-def}
\end{equation}
When $\tau\gg\tau_{c}$, $P_{S}$ is comparable to $P_{S}^{\rm{max}}(\tau\rightarrow\infty)$, and a pulse width of about $\tau_{c}$ provides a sufficiently high $P_{S}$ as shown in Fig.~\ref{figPs}.


\begin{figure*}
	\includegraphics[clip,width=16cm]{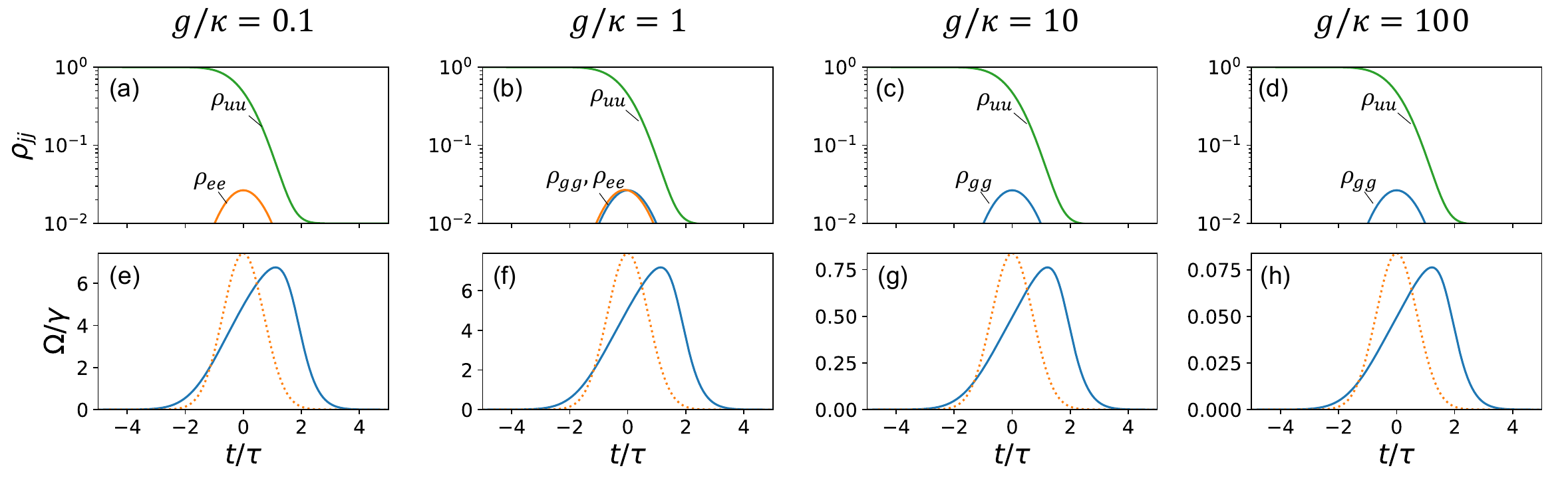}
	\caption{Temporal analysis at $C=10$, $\eta_{\rm{esc}}=0.95$, and $P_{S}=0.99P_{S}^{\rm{max}}(\tau\rightarrow\infty)$ of the adiabatic condition corresponding to the circle marks ($\bigcirc$) indicated in Fig. 2.
	(a)--(d) are the population dynamics of $\rho_{uu}$ (green), $\rho_{ee}$ (orange) and $\rho_{gg}$ (blue).
	(e)--(h) show $\Omega(t)$ (blue) and $w_{0}(t)$ (orange dotted) for reference.
	(a,e) $g/\kappa=0.1, (g/\gamma,\kappa/\gamma)=(200,2000), \tau=10\tau_{c}$, corresponding to the Purcell regime denoted as A in Fig. 2.
	(b,f) $g/\kappa=1, (g/\gamma,\kappa/\gamma)=(20,20), \tau=10\tau_{c}$, corresponding to the intermediate regime denoted as B in Fig. 2.
	(c,g) $g/\kappa=10, (g/\gamma,\kappa/\gamma)=(2,0.2), \tau=10\tau_{c}$, corresponding to the strong-coupling regime denoted as C in Fig. 2.
	(d,h) $g/\kappa=100, (g/\gamma,\kappa/\gamma)=(0.2,0.002), \tau=10\tau_{c}$, corresponding to the weak-coupling regime under the condition of $C>1$, denoted as D in Fig. 2. We perform the numerical calculations at the range of $-5\tau<t<+5\tau$}.
	\label{fig3}
\end{figure*}

\begin{figure*}
	\includegraphics[clip,width=16cm]{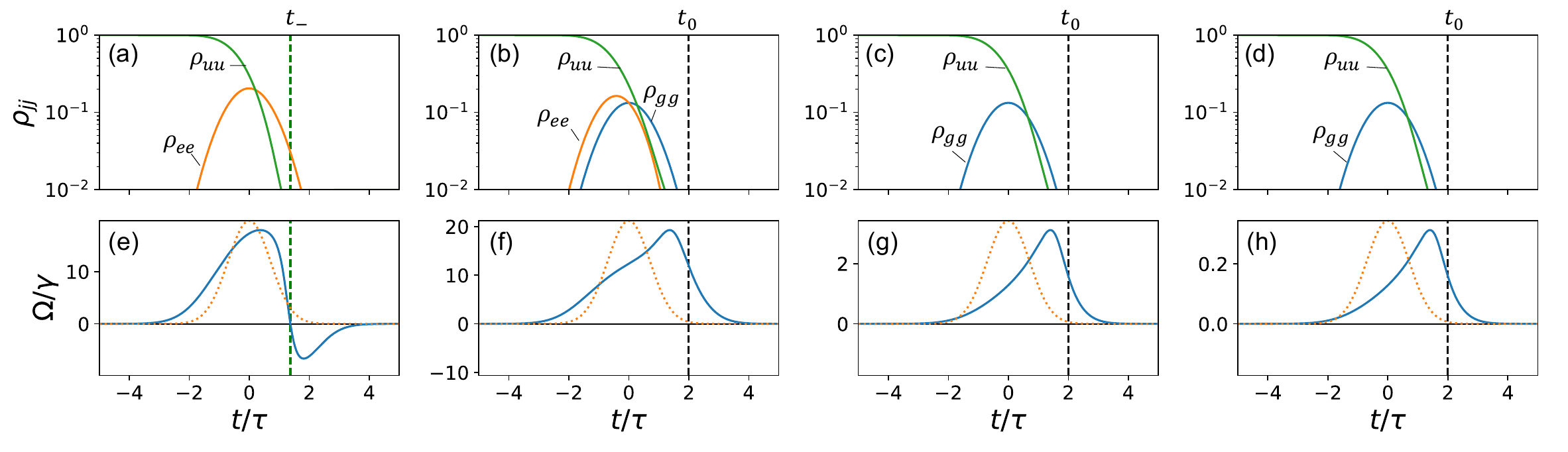}
	\caption{Temporal analysis at $C=10$, $\eta_{\rm{esc}}=0.95$, and $P_{S}=0.99P_{S}^{\rm{max}}(\tau\rightarrow\infty)$ in the nonadiabatic condition corresponding to the star marks ($\star$) shown in Fig. 2.
	The lines and conditions are the same as Fig. 3 except for $\tau$: (a,e) are $\tau=1.3\tau_{c}$, and (b--d) and (f--h) are all $\tau=2\tau_{c}$.}
	\label{fig4}
\end{figure*}

\begin{figure*}
	\includegraphics[clip,width=16cm]{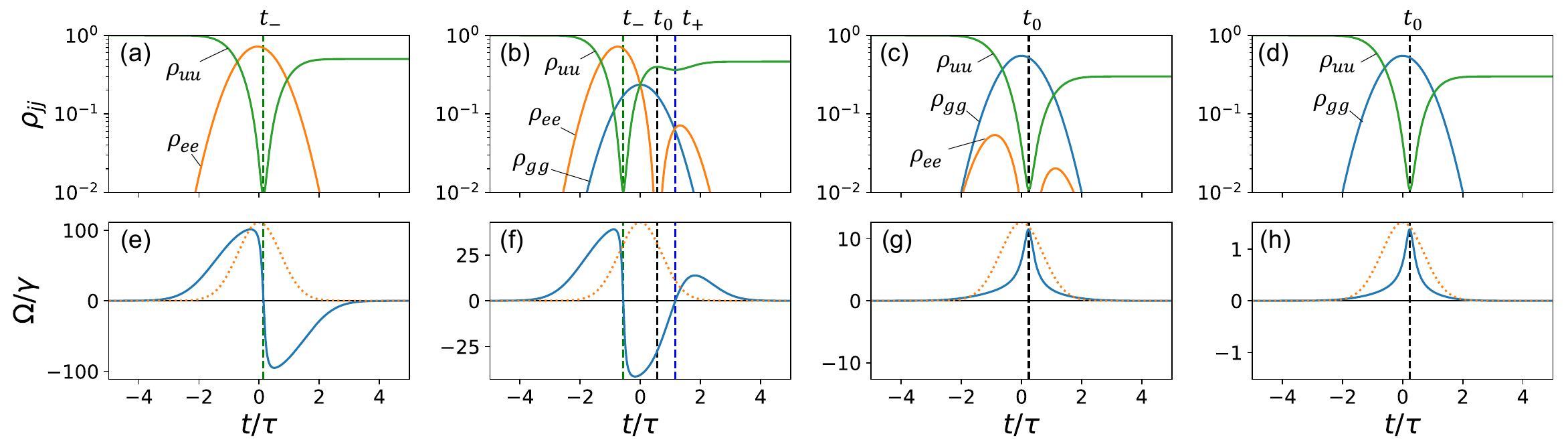}
	\caption{Temporal analysis at $C=10$, $\eta_{\rm{esc}}=0.95$, and $P_{S}=0.5P_{S}^{\rm{max}}(\tau\rightarrow\infty)$ in the nonadiabatic condition corresponding to the cross marks ($\times$) shown in Fig. 2.
	The lines and conditions are the same as Fig. 3 except for $\tau$: (a,e) are $\tau=0.19\tau_{c}$, (b,f) are $\tau=0.57\tau_{c}$, and (c,d) and (g,h) are all $\tau=0.24\tau_{c}$}
	\label{fig5}
\end{figure*}

\section{\label{sec3}Temporal analysis}
Here, we investigate the population dynamics and required external fields in all coupling regimes. We first derive the necessary temporal shape of $\Omega(t)$ for the Gaussian wavepacket emission in the resonant case (${\Delta_{u}=\Delta_{e}=0}$).
Then, $\Omega(t)$ is expressed using Eq.~(\ref{eq-alpha-u}) as~\cite{vasilev2010single}
\begin{align}
\Omega(t) 
&= -\frac{\dot{\rho_{uu}}}{2\alpha_{e}\sqrt{\rho_{uu}}}\notag\\
&= \frac{\sqrt{P_{S}}e^{-\frac{t^{2}}{2\tau^{2}}}}{g\tau^{9/2}\sqrt{2\kappa_{\rm{ex}}\sqrt{\pi}}}\frac{(t-t_{+})(t-t_{-})}{\sqrt{\rho_{uu}}},
\label{omegat}
\end{align}
where a global phase is neglected. Practically, $\Omega(t)$ can be controlled by modulating the amplitude and phase of the external input light. 
We note that in the nonresonant case (${\Delta_{u},\Delta_{e}\neq0}$), we also derive the expression of $\Omega(t)$, as shown in Appendix~A.

Figures~\ref{fig3}--\ref{fig5} show representative population dynamics and $\Omega(t)$, derived from Eqs.~(\ref{eq-rhogg})--(\ref{eq-rhouu}) and Eq.~(\ref{omegat}), respectively. Here, we set $C=10$ and $\eta_{\rm{esc}}=0.95$. 
Figure~\ref{fig3} shows the dynamics of the adiabatic condition, where $P_{S}=0.99P_{S}^{\rm{max}}(\tau\rightarrow\infty)$ and $\tau=10\tau_{c}$. Figures~\ref{fig4} and \ref{fig5} show the dynamics of the nonadiabatic condition, where $P_{S}=0.99P_{S}^{\rm{max}}(\tau\rightarrow\infty)$ and $P_{S}=0.5P_{S}^{\rm{max}}(\tau\rightarrow\infty)$, respectively.

Figures~\ref{fig3}(a,e), (b,f), (c,g), and (d,h) show the population dynamics and external fields in the Purcell, intermediate, strong-, and weak-coupling regimes in the adiabatic condition, respectively. For the adiabatic condition of $\tau\gg\tau_{c}$, there is only $10^{-2}$ order difference of $\rho_{gg}$ and $\rho_{ee}$ among the coupling regimes, limited by our choice of $P_{S}=0.99P_{S}^{\rm{max}}(\tau\rightarrow\infty)$. The required $\Omega(t)$ shape is also similar. 

In contrast, as shown in Fig.~\ref{fig4}, for the nonadiabatic condition of $\tau\simeq\tau_{c}$, relatively large differences of $\rho_{gg}$ and $\rho_{ee}$ arise. The atomic excitation is not negligible in the Purcell regime and is negligible in the strong- and weak-coupling regimes. The crossover of these two regimes is seen as the intermediate regime.

In the Purcell regime shown in Fig.~\ref{fig4}(e), there are two parts to $\Omega(t)$: $\Omega(t)>0$ is a driving pulse to excite the atom, analogous to a $\pi$ pulse for $\rho_{uu}\rightarrow0$, and $\Omega(t)<0$ brings the atom back to the $|u\rangle$ state by flipping the phase in order to keep the Gaussian shape of the photon wavepacket. This is closely related to single-photon generation for a two-level atom, which emits a non-Gaussian shape wavepacket~\cite{cui2005quantum}, by applying a fast $\pi$ pulse and observing the photon emitted into the cavity mode. The difference between the two- and three-level schemes is the existence of the uncoupled $|u\rangle$ state, which allows us to control the output pulse shape.

On the other hand, in the strong- and weak-coupling regimes shown in Fig.~\ref{fig4}(g,h), the magnitude of $\Omega$ must be increased adiabatically and eventually brought to $\Omega>g$ to minimize the excitation to $|e\rangle$. The absence of $\rho_{ee}$ manifests due to the adiabatic transfer of the population from $\rho_{uu}$ to $\rho_{gg}$. A photon is generated by vSTIRAP, confined in the cavity, and then emitted through the output coupler. We note that in the weak-coupling regime, where $\gamma$ is larger than $g$ and $\kappa$, the picture of vSTIRAP is valid so long as there is no excitation. If $C$ is sufficiently large, vSTIRAP can be used to emit a single photon with a high success probability under a resonant condition as well as a nonresonant condition in the weak-coupling regime, which is realized in ion-trap systems~\cite{ion1,ion2}.
Thus, single-photon generation in the strong- and weak-coupling regimes is a different process to that in the Purcell regime, even though they have the same $C$ and achieve the same $P_{S}$.
Interestingly, Fig.~\ref{fig4}(b) shows the dynamics in the intermediate regime, which can be seen in the crossover between (a) and (c). Therefore, the excitation of the atom and adiabatic transfer occur simultaneously.

It is possible to generate shorter pulse widths than $\tau_{c}$ by sacrificing $P_{S}$.
In this case, we can generate Gaussian shaped photons with arbitrary pulse width by transferring the population back from $|g\rangle$ to $|u\rangle$. 
There is then some remaining population in $|u\rangle$ at the end of single-photon generation. This corresponds to the case of $\rho_{uu}(t=\infty)\neq0$ in Eq~(\ref{casei}).
Figure~\ref{fig5} shows an example of generating short pulses by sacrificing $P_{S}$, where $P_{S}=0.5P_{S}^{\rm{max}}(\tau\rightarrow\infty)$.
In the Purcell regime shown in Fig.~\ref{fig5}(a,e), the driving pulse for $\rho_{gg}\rightarrow\rho_{uu}$ needs a larger pulse area than the case shown in Fig.~\ref{fig4}(a,e), in order to maintain the larger $\rho_{uu}$ at the end of single-photon generation.
In the strong- and weak-coupling regimes shown in Fig.~\ref{fig5}(c,g) and (d,h), the $\rho_{gg}\rightarrow\rho_{uu}$ driving is realized by vSTIRAP, i.e., by adiabatically modulating  $\Omega(t)$ from ${\Omega>g}$ to ${\Omega<g}$.
We can also observe oscillatory dynamics in the intermediate regime as shown in Fig.~\ref{fig5}(b,f).

\section{\label{sec4}External loss rate optimization}
We optimize $\kappa_{\rm{ex}}$ in order to achieve high $P_{S}$ and short $\tau$ simultaneously using Eq.~(\ref{Ps-gen}).
Here, we introduce an {\it internal cooperativity}~\cite{goto2019figure}, $C_{\rm{in}}\equiv g^{2}/(2\kappa_{\rm{in}}\gamma)$, which is a cooperativity parameter with respect to $\kappa_{\rm{in}}$ rather than $\kappa$ in the case of the standard cooperativity, where ${C=C_{\rm{in}}(1-\eta_{\rm{esc}})}$.
In the adiabatic limit, by optimizing $\kappa_{\rm{ex}}$, we can obtain the upper bound of the overall success probability, $P_{S}^{\rm{ub,opt}}$, using $C_{\rm{in}}$ by setting $\kappa_{\rm{ex}}=\kappa_{\rm{ex}}^{\rm{ub,opt}}$ as~\cite{goto2019figure}
\begin{gather}
P_{S}^{\rm{ub,opt}} = 1-\frac{2}{1+\sqrt{2C_{\rm{in}}+1}}
\approx 1-\sqrt{\frac{2}{C_{\rm{in}}}},
\label{eq-Psopt}\\
\kappa_{\rm{ex}}^{\rm{ub,opt}} = \kappa_{\rm{in}}\sqrt{2C_{\rm{in}}+1}.
\label{eq-kexopt}
\end{gather}
This approximation holds when $C_{\rm{in}}\gg1$.
We now numerically investigate $P_{S}^{\rm{max}}$ and the optimal $\kappa_{\rm{ex}}$ with respect to various $\tau$, including nonadiabatic conditions, using Eqs. (\ref{eq-t0}-\ref{Ps-gen}). 

\begin{figure*}
	\includegraphics[clip,width=15cm]{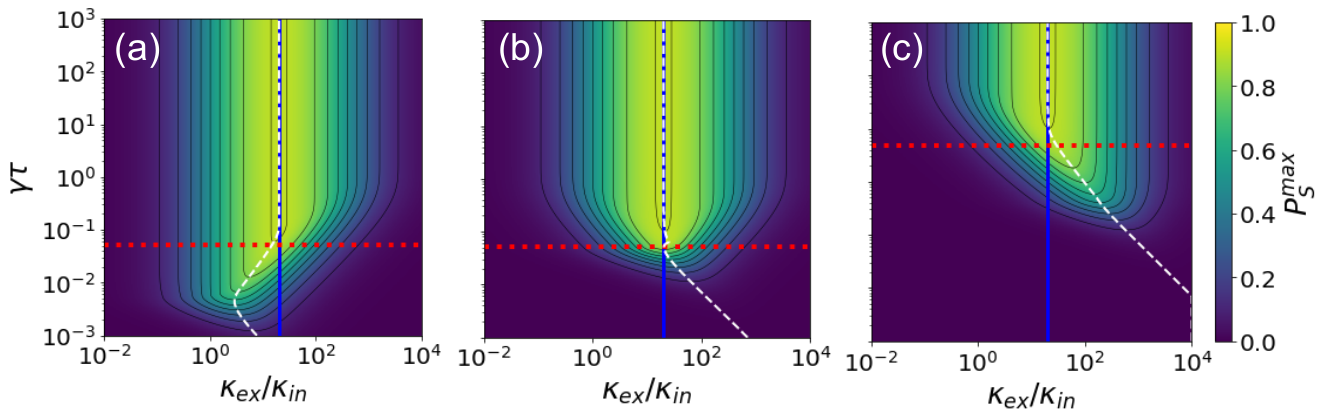}
	\caption{Cavity external loss optimization for $\kappa_{\rm{ex}}/\kappa_{\rm{in}}$, for each $\tau$, where $C_{\rm{in}}=200$, (a) $\kappa_{\rm{in}}/\gamma = 100$, (b) $1$ and (c) $0.01$. The white dashed, red dotted and blue lines represent the maximum $P_{S}$ in each $\tau$, $\tau_{c}= \rm{max}(1/\kappa^{\rm{ub,opt}},\kappa^{\rm{ub,opt}}/g^{2})$, and $\kappa_{\rm{ex}}/\kappa_{\rm{in}}=\sqrt{2C_{\rm{in}}+1}$, respectively.}
	\label{fig6}
\end{figure*}

Figure \ref{fig6} shows $P_{S}^{\rm{max}}$ for $\tau$ and $\kappa_{\rm{ex}}$ with ${C_{\rm{in}}=200}$ and ${\kappa_{\rm{in}}/\gamma=100,1,0.01}$.
Note that these parameters are experimentally accessible using, for instance, nanofiber cavities~\cite{sam2020}.
Here, the critical width of the photon is determined by $\tau_{c}$ in Eq.~(\ref{tauc-def}) (red dotted lines), where ${\kappa=\kappa^{\rm{ub,opt}} \equiv \kappa_{\rm{in}}(1+\sqrt{2C_{\rm{in}}+1})}$ is used.
This shows that the optimal $\kappa_{\rm{ex}}$ (white dashed lines) becomes $\kappa_{\rm{ex}}^{\rm{ub,opt}}$ (blue lines) in the $\tau\gtrsim\tau_{c}$ region.
On the other hand, as the pulse width decreases to $\tau\lesssim\tau_{c}$, the optimal $\kappa_{\rm{ex}}$ deviates from  $\kappa_{\rm{ex}}^{\rm{ub,opt}}$ in order to satisfy the bandwidth of the dressed atom for short pulses. 
From the relation ${\kappa_{\rm{in}}/\gamma\approx(\kappa/g)^{2}}$ when ${\sqrt{C_{\rm{in}}}\gg1}$ and ${\kappa\approx\kappa^{\rm{ub,opt}}}$~\cite{memo3}, the deviation has different characteristics with respect to the coupling regimes. 
In the Purcell regime, as $\tau_{c}$ is defined as $\kappa/g^{2}$, $\tau$ is shortened by decreasing $\kappa$. 
On the other hand, in the strong-coupling regime, $\tau_{c}$ is defined as $1/\kappa$, hence $\tau$ is shortened by increasing $\kappa$, which is captured by the sign of $d\tau_{c}/d\kappa$. 
Thus, for short pulse generation, one needs to account for the shifted optimal $\kappa_{\rm{ex}}$ from $\kappa_{\rm{ex}}^{\rm{ub,opt}}$.

\begin{figure*}
	\includegraphics[clip,width=14cm]{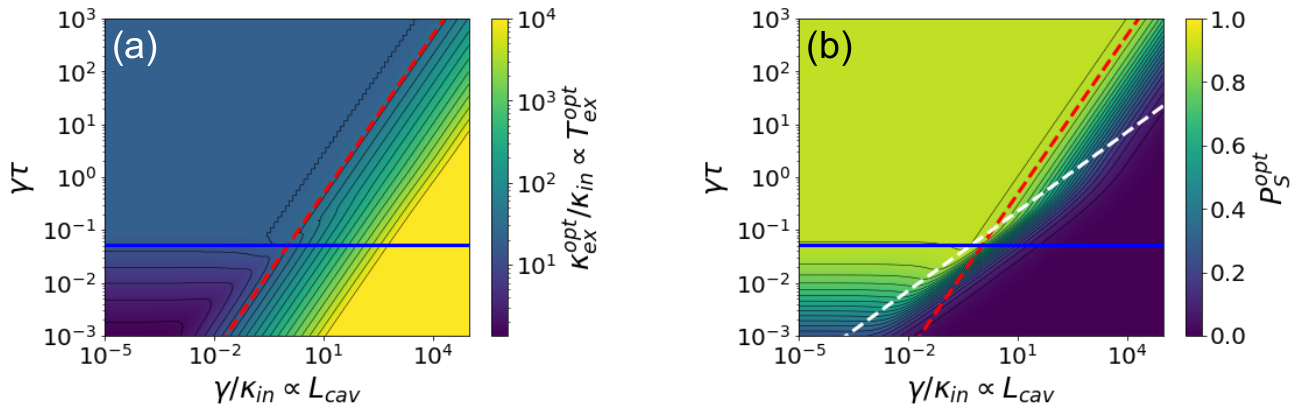}
	\caption{$\tau$ and $L_{\rm{cav}}$ dependencies of (a) $T_{\rm{ex}}^{\rm{opt}}$ and (b) $P_{S}^{\rm{opt}}$ after cavity transmittance optimizations with $C_{\rm{in}}=200$. The red dashed and blue lines represent $1/\kappa^{\rm{ub,opt}}$ and $\kappa^{\rm{ub,opt}}/g^{2}$, respectively. The white dashed line represents $1/g$.}
	\label{fig7}
\end{figure*}

\section{\label{sec5}Example}
Using the analytical results derived above, we provide an example of how to achieve a high success probability while maintaining a short pulse width.
To explore the dependence on cavity QED parameteres, we explicitly write $(g, \kappa_\mathrm{in}, \kappa_\mathrm{ex})$ in natural units ($c=1$) as
 ${g=\sqrt{{\gamma}/({2\tilde{A}_{\rm{eff}}L_{\rm{cav}}})}}$, ${\kappa_{\rm{in}}=\alpha_{\rm{loss}}/(4L_{\rm{cav}})}$, and ${\kappa_{\rm{ex}}=T_{\rm{ex}}/(4L_{\rm{cav}})}$~\cite{reiserer2015cavity}. Here, $\tilde{A}_\mathrm{eff}$ is the effective cross-sectional area of the cavity mode at the atomic position ($\tilde{A}_\mathrm{eff}\equiv A_\mathrm{eff}/\sigma_0$, where $\sigma_0$ is the on-resonant scattering cross section of atom), $L_\mathrm{cav}$ is the cavity length, $\alpha_\mathrm{loss}$ is the round-trip cavity internal loss, and $T_\mathrm{ex}$ is the transmittance of the output coupler of the cavity.
The internal cooperativity and the ratio of external and internal loss rates are given by  ${C_{\rm{in}}=1/(\tilde{A}_{\mathrm{eff}}\alpha_{\mathrm{loss}})}$ and $\kappa_\mathrm{ex}/\kappa_\mathrm{in}=T_\mathrm{ex}/\alpha_\mathrm{loss}$, respectively.

As an example, we consider an experimental setup consisting of an optical nanofiber cavity~\cite{sam2020}, where both $\alpha_{\rm{loss}}$ and $\tilde{A}_{\rm{eff}}$ are independent of $L_{\rm{cav}}$.
When $C_\mathrm{in}\gg1$, the condition for high $P_S^\mathrm{max}$ in Eq.~(\ref{tauccase}) when $\tau\gg\tau_{c}$ can be rewritten as
\begin{equation}
\tau_{c} =
\left\{
\begin{array}{ll}
	4L_\mathrm{cav}/(T_\mathrm{ex}+\alpha_\mathrm{loss}) & \text{if } g>(\kappa,\gamma) \text{ or } \gamma>g>\kappa,\\
	\tilde{A}_\mathrm{eff}(T_\mathrm{ex}+\alpha_\mathrm{loss})/(2\gamma) & \text{if } \kappa>g^{2}/\kappa>\gamma.
\end{array}
\label{tauccase2}
\right.
\end{equation}
Here, $\tau_c$ is proportional to the cavity length in the strong- and weak-coupling regimes, while $\tau_c$ is independent of the cavity length in the Purcell regime.
For given cavity parameters ($\tilde{A}_\mathrm{eff}$, $T_\mathrm{ex}, \alpha_\mathrm{loss}$, $\gamma$), the shortest $\tau_c$ can be realized in the Purcell regime by sufficiently shortening the cavity length $L_\mathrm{cav}$.

We can further optimize the external coupling rate or cavity transmittance $\kappa_\mathrm{ex}/\kappa_{\rm{in}}\propto T_\mathrm{ex}$ to find the highest success probability $P_S^\mathrm{opt}$ for a given ($\tilde{A}_\mathrm{eff}$, $\alpha_\mathrm{loss}$) or fixed $C_\mathrm{in}$. Figure~\ref{fig7}(a) and \ref{fig7}(b) illustrate numerically optimized external coupling rate or cavity transmittance $\kappa_\mathrm{ex}^\mathrm{opt}/\kappa_\mathrm{in}\propto T_{\rm{ex}}^{\rm{opt}}$ and $P_S^\mathrm{opt}$, respectively, as a function of $\tau$ and $L_\mathrm{cav}$.
We note that Fig.~\ref{fig7} shows the cavity-length dependence via $\gamma/\kappa_\mathrm{in}\propto L_\mathrm{cav}$ along the horizontal axis, while the related plot in Fig.~2 shows the dependence on ${g/\kappa\propto \sqrt{L_{\rm{cav}}}}$.
Figure~\ref{fig7}(a) reconfirms that $\kappa_\mathrm{ex}^\mathrm{opt}$ is almost constant at $\kappa_\mathrm{ex}^\mathrm{ub,opt}$ of Eq.~(\ref{eq-kexopt}) for $\tau\gtrsim\tau_{c}$, while $\kappa_\mathrm{ex}^\mathrm{opt}$ widely varies with $L_\mathrm{cav}$ for $\tau\lesssim\tau_{c}$.
In addition, $P_{S}^\mathrm{opt}$ shown in Fig.~\ref{fig7}(b) is close to the upper bound $P_{S}^{\rm{ub,opt}}$ for ${\tau\gtrsim\tau_{c}}$, while it rapidly decreases for $\tau\lesssim\tau_{c}$. We note that in contrast to that shown in Fig.~2, the dependence on $1/g$ can be seen, i.e., $P_{S}^{\rm{opt}}$ rapidly decreases below the $1/g$ line.
This suggests that the adiabaticity constraint of vSTIRAP, which is referred to as a first lower limit~\cite{kuhn1999controlled}, is shown as a result of optimizing $\kappa_{\rm{ex}}$ to bring $\kappa$ closer to $g$.
These observations again show that the shortest $\tau$ can be achieved in the Purcell regime with sufficiently short $L_\mathrm{cav}$. 
Overall, $\tau_{c}$ is a good indicator for the tradeoff between short pulses and high success probability.

The approach for achieving a high success probability with a short pulse width can be summarized into three conditions when $C_\mathrm{in}\gg1$: (i) choosing the optimized cavity transmittance, i.e., $T_\mathrm{ex}\approx \sqrt{2\alpha_\mathrm{loss}/\tilde{A}_\mathrm{eff}}$, (ii) shortening the cavity length $L_\mathrm{cav}$ in the Purcell regime, i.e., $L_\mathrm{cav}\lesssim \alpha_\mathrm{loss}/(4\gamma)$, and (iii) choosing the pulse width longer than the critical pulse width, i.e., $\tau>\tau_c$.

\section{\label{summary}summary}
In this paper, we have investigated single-photon generation using a three-level $\Lambda$-type atom including nonadiabatic conditions, by considering a model which emits a Gaussian wavepacket. 
We have derived the analytical expression for the success probability for arbitrary coupling regime and pulse width. We have analyzed the temporal dynamics in different coupling regimes, and have provided a method for setting the cavity QED parameters and pulse width in order to achieve a high success probability while keeping a short pulse width. In addition, we have shown the correction method of the arbitrary detuning by optimally controlling the external field.
This model is convenient for comprehensively analyzing temporal dynamics with arbitrary coupling condition.
These results are also applicable to single-photon storage, which is the time-reversal situation. Furthermore, in the non-Gaussian waveform cases, similar arguments will hold. Among the non-Gaussian shapes, we believe that the unimodal function, which is a real function having only one maximum, can be solved at least numerically, and the same relationship between the generation efficiency and photon width as in Gaussian can be obtained.
\section*{ACKNOWLEDGMENTS}
The authors thank Akinori Suenaga, Daiki Kojima, Rui Asaoka, Rina Kanamoto, Samuel Ruddell, and Karen Webb for their useful comments. This work was supported by JST CREST, Grant Number JPMJCR1771, and JST Moonshot R$\&$D, Grant Number JPMJMS2061, Japan.
\\

\section*{APPENDIX A: Derivation of $\Omega(t)$ for $\Delta_{u},\Delta_{e}\neq0$} 
We discuss the external control field $\Omega(t)$ for non-zero time-independent detunings of $\Delta_u$ and $\Delta_e$, and an arbitrary waveform of output pulse $w_{0}(t)\in \mathbb{C}$.
The difference of $\Omega(t)$ from the resonant case of $\Delta_u=\Delta_e=0$ lies in the fact that $\alpha_u$ and $\Omega(t)$ take on complex values as seen in Eqs. (\ref{eq-alpha-u}) and (\ref{eq-alpha-e}).
Thus, we rewrite $\alpha_u(t)$ and $\Omega(t)$ in the exponential form of
\begin{align}
\Omega(t)=\Omega_{0}(t)e^{i\phi_{0}(t)},~\alpha_{u}(t)=\sqrt{\rho_{uu}(t)}e^{i\phi_{u}(t)},\tag{A1}
\end{align}
where $\Omega_{0}(t),\phi_{0}(t),\phi_{u}(t)\in\mathbb{R}$.
In the following, we outline the derivation of $\Omega_{0}(t)$, $\phi_{0}(t)$, and $\phi_{u}(t)$ by using Eqs. (\ref{eq-alpha-u}) and (\ref{eq-alpha-e}).

First, to simplify the expression, we introduce the complex parameter $z$ as
\begin{equation}
z\equiv \dot{\alpha}_e+(\gamma+i\Delta_e)\alpha_e -g\alpha_g,\tag{A2}
\end{equation}
where $z$ is uniquely determined by $w_{0}(t)$ and the atom-cavity parameters.
Eq. (\ref{eq-alpha-e}) can be rewritten in terms of $z$ as
\begin{equation}
z=\Omega(t)\alpha_u(t).\tag{A3}
\end{equation}
We then obtain $\Omega_0(t)$ and the phase relation between $\phi_0(t)$ and $\phi_u(t)$ as
\begin{align}
\Omega_0(t)&=\frac{|z|}{\sqrt{\rho_{uu}(t)}},\tag{A4}\\
\phi_0(t)&=-\phi_u(t)-\frac{i}{2}\mathrm{ln}\left(\frac{z}{z^*}\right).\tag{A5}
\end{align}

Next, we derive the explicit expression of $\phi_u(t)$ by using Eq.~(\ref{eq-alpha-u}).
By substituting the expression of $\dot{\alpha}_{u}$ obtained from Eq. (A1), given by
\begin{equation}
\dot{\alpha}_u=\alpha_u\left[\frac{\dot{\rho}_{uu}(t)}{2\rho_{uu}(t)}+i\dot{\phi}_{u}(t)\right],\tag{A6}
\end{equation}
into Eq. (\ref{eq-alpha-u}), we obtain
\begin{gather}
\Omega^{*}(t)
=-\frac{\alpha_{u}(t)}{\alpha_{e}(t)}\left[y+i\dot{\phi}_{u}(t)\right],\tag{A7}
\end{gather}
where $y\equiv i\Delta_{u}+\frac{\dot{\rho}_{uu}(t)}{2\rho_{uu}(t)}$ has been introduced.
Substituting Eq.~(A7) into $z/z^{*}$, we finally obtain
\begin{gather}
\dot{\phi}_{u}(t)=-\frac{\mathrm{Im}\left[\alpha^{*}_{e}yz\right]}{\mathrm{Re}\left[\alpha^{*}_{e}z\right]}
\rightarrow
\phi_{u}(t)=-\int_{t_{\rm{st}}}^{t}\frac{\mathrm{Im}\left[\alpha^{*}_{e}yz\right]}{\mathrm{Re}\left[\alpha^{*}_{e}z\right]}dt'.\tag{A8}
\end{gather}
Equations (A4), (A5), and (A8) provide an explicit form of $\Omega(t)$ for $\Delta_u, \Delta_e\neq 0$, which enables us to reach the upper bound of success probability $P_S^\mathrm{max}$ in Eq. (\ref{Ps-gen}).

In addition, in the case of $w_{0}(t)\in \mathbb{R}$, since ${\alpha_{g}, \alpha_{e}\in\mathbb{R}}$ holds from Eqs.~(\ref{eq-alpha-e}), (\ref{eq-alpha-g}), and (\ref{eq-w0}), 
$\dot{\rho}_{uu}(t)$ is rewritten by using Eqs.~(\ref{eq-alpha-e}), (\ref{eq-alpha-g}), and (\ref{eq-ro1}) as
\begin{gather}
\dot{\rho}_{uu}(t)=2\alpha_{e}(t)\left[i\Delta_{e}\alpha_{e}(t)-z\right],\tag{A9}\\
\rightarrow z=i\Delta_{e}\alpha_{e}(t)-\frac{\dot{\rho}_{uu}(t)}{2\alpha_{e}(t)}.\tag{A10}
\end{gather}
Thus, $\phi_u(t)$ is given by
\begin{gather}
\phi_{u}(t)=-\Delta_{u}(t-t_\mathrm{st})+\Delta_{e}\int_{t_{\rm{st}}}^{t}\frac{\alpha_{e}^{2}(t')}{\rho_{uu}(t')}dt'.\tag{A11}
\end{gather}

Note that similar results under the $\kappa\gg g$ condition are obtained in Refs.~\cite{Giannelli,Morin}.

\section*{APPENDIX B: Derivation of $P_{S}$ for finite $\tau$ and ${t\rightarrow\infty}$} 
Using $g=\sqrt{2C\kappa\gamma}$, Eq.~(\ref{eq-rhouu}) can be rewritten when ${t\rightarrow\infty}$ as
\begin{align}
&\rho_{uu}(t) 
= 1-\frac{P_{S}\left[\gamma+2\kappa\tau^{2}(g^{2}+\gamma\kappa)\right]\left[\mathrm{Erf}\left(\frac{t}{\tau}\right)+1\right]}{4g^{2}\kappa_{\rm{ex}}\tau^{2}}\notag\\
&-\frac{P_{S}e^{ - \frac{t^{2}}{\tau^{2}}}}{2\sqrt{\pi}\kappa_{\rm{ex}}g^{2}\tau^{5}}\left\{t^{2}-t(\gamma+2\kappa)\tau^{2}+\left[g^{2}+\kappa(2\gamma+\kappa)\right]\tau^{4}\right\}\notag\\
&\xrightarrow{t\rightarrow\infty} 1-\frac{2P_{S}\left[\gamma+2\kappa\tau^{2}(g^{2}+\gamma\kappa)\right]}{4g^{2}\kappa_{\rm{ex}}\tau^{2}}\notag\\
&=1-\frac{P_{S}\left(1/\kappa+4C\kappa\tau^{2}+2\kappa\tau^{2}\right)}{4C\kappa_{\rm{ex}}\tau^{2}}.\tag{B1}\label{eq-rhouu-ap}
\end{align}
Solving Eq.~(\ref{eq-rhouu-ap}) for $P_{S}$ leads 
\begin{align}
P_{S}&= \left[1-\rho_{uu}(t=\infty)\right]\frac{2C\kappa_{\rm{ex}}\tau^{2}}{\frac{1}{2\kappa}+\left(2C+1\right)\kappa\tau^{2}}.\tag{B2}
\label{P_S_t_infty}
\end{align}
For $\tau\gg\frac{1}{\kappa\sqrt{2(2C+1)}}$, Eq.~(\ref{P_S_t_infty}) can be approximated as
\begin{align}
P_{S} &\approx \eta_{\rm{esc}}\left[1-\rho_{uu}(t=\infty)\right]\frac{2C}{2C+1}.\tag{B3}
\end{align}
On the other hand, in the limit of $C\rightarrow\infty$, Eq.~(\ref{P_S_t_infty}) becomes
\begin{align}
P_{S}&\xrightarrow{C\rightarrow\infty}\frac{\kappa_{\rm{ex}}}{\kappa}[1-\rho_{uu}(t=\infty)].\tag{B4}\label{eq-b2}
\end{align}

\section*{APPENDIX C: The effect of the truncation in Gaussian pulse} 
\begin{figure}
	\includegraphics[clip,width=9cm]{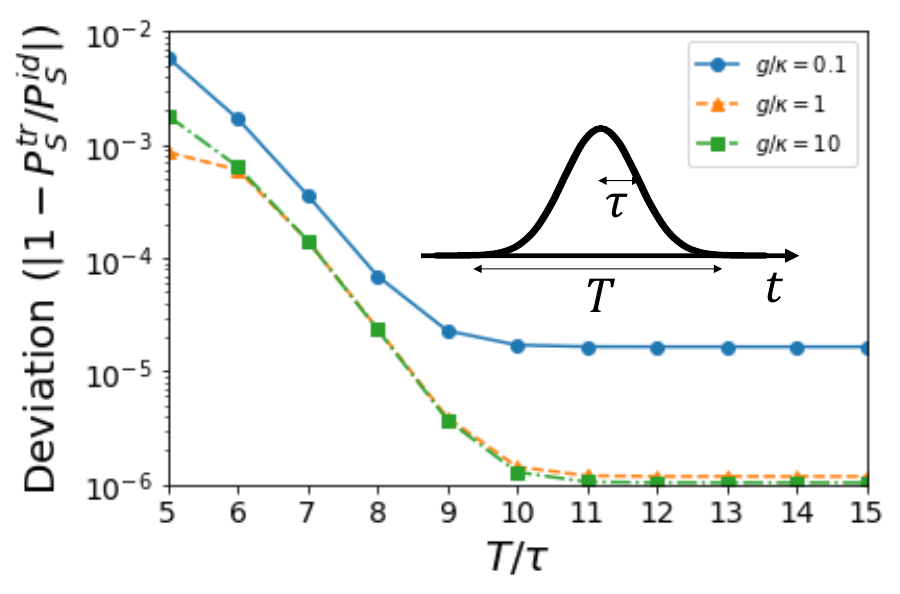}
	\caption{Simulation results of the deviation due to the truncation for truncated Gaussian pulse, calculated with QuTiP~\cite{Qutip1,Qutip2}, for $g/\kappa=0.1$ (blue solid line with circle), $g/\kappa=1$ (orange dashed line with triangle), and $g/\kappa=10$ (green dash-dot line with square), where $C=10$, $\eta_{\rm{esc}}=0.95$, $P_{S}=0.99P_{S}^{\rm{max}}(\tau\rightarrow\infty)$, and $\tau=2\tau_c$. The error of the numerical simulation is limited to about $10^{-5}$ for $g/\kappa=0.1$ and $10^{-6}$ for $g/\kappa=1,10$. The range of numerical integration is $15\tau$ at all plot points.
}
	\label{fig-trunc}
\end{figure}
   We show that the results of this paper are also useful for truncated pulses. We define the truncated Gaussian pulse as 
\begin{equation}
w_0(t) = 
\left\{
\begin{array}{ll}
	\mathcal{N} \left[\exp\left( - \frac{t^{2}}{2\tau^{2}}\right)-\exp\left( - \frac{T^{2}}{8\tau^{2}}\right)\right],
	&|t|<T/2,\\
	0, &\mathrm{otherwise},
\end{array}
\right.\tag{C1}
\label{Trunc-Gaussian}
\end{equation}
where $\mathcal{N}$ denotes the normalized constant, and the photon emission starts at ${t_{\rm{st}}=-T/2}$ and ends at $t_{\rm{end}}=T/2$.
Substituting this truncated pulse in place of Eq.~(\ref{Gaussian}) to obtain $\Omega(t)$ numerically as in Eq.~(\ref{omegat})~\cite{vasilev2010single}. We denote the photon generation efficiency using Eq.~(\ref{Trunc-Gaussian}) as $P_{S}^{\rm{tr}}$. 
In contrast, we denote the pre-defined target generation efficiency as $P_{S}^{\rm{id}}$. We then define the deviation due to the truncation as $1-P_{S}^{\rm{tr}}/P_{S}^{\rm{id}}$. The simulation results of the relationship between this deviation and the truncated length $T/\tau$ are shown in Fig. 8.
We see that in $T/\tau>6$, the generation efficiency can be calculated with an deviation of approximately 0.1\%. The numerical simulation of the single photon generation is performed with an open-source software QuTiP~\cite{Qutip1,Qutip2}. Note that the error of the numerical simulation is limited to about $10^{-5}$ for $g/\kappa=0.1$ and $10^{-6}$ for $g/\kappa=1,10$.


\begin{thebibliography}{10}
\bibitem{memo}
Substituting $g=\frac{2C\gamma}{g/\kappa}$ and $\kappa=\frac{2C\gamma}{(g/\kappa)^{2}}$ into Eqs.~(\ref{eq-t0}--\ref{Ps-gen}), $P_{S}^{\rm{max}}$ can be rewritten using ${C,~\eta_{\rm{esc}},~g/\kappa}$ and $\gamma\tau$.
\bibitem{memo3}
When ${\sqrt{C_{\rm{in}}}=\sqrt{\frac{g^{2}}{2\kappa_{\rm{in}}\gamma}}\gg1}$ and \\
${\kappa\approx\kappa^{\rm{ub,opt}}=\kappa_{\rm{in}}\left(1+\sqrt{2C_{\rm{in}}+1}\right)}$,
${\frac{g}{\kappa}\approx \frac{g}{\kappa^{\rm{ub,opt}}}=\frac{\sqrt{2C_{\rm{in}}\kappa_{\rm{in}}\gamma}}{\kappa_{\rm{in}}\left(1+\sqrt{2C_{\rm{in}}+1}\right)}\approx\frac{\sqrt{2C_{\rm{in}}\kappa_{\rm{in}}\gamma}}{\kappa_{\rm{in}}\sqrt{2C_{\rm{in}}}}=\sqrt{\frac{\gamma}{\kappa_{\rm{in}}}}}$\\ is derived.
\end{thebibliography}
\end{document}